**Physics-Driven Self-Supervised Deep Learning for Free-Surface Multiple Elimination**

Jing Sun[1*], Tiexing Wang[2], Eric Verschuur[3], Ivan Vasconcelos[2]

*1. Faculty of Electrical Engineering, Mathematics and Computer Science, Delft University of Technology, The Netherlands (Jing.Sun@tudelft.nl)*
*2. R&D Department, Shearwater GeoServices, UK*
*3. Faculty Of Civil Engineering and Geosciences, Delft University of Technology, The Netherlands*

**Summary**

In recent years, deep learning (DL) has emerged as a promising alternative approach for various seismic processing tasks, including primary estimation (or multiple elimination), a crucial step for accurate subsurface imaging. In geophysics, DL methods are commonly based on supervised learning from large amounts of high-quality labelled data. Instead of relying on traditional supervised learning, in the context of free-surface multiple elimination, we propose a method in which the DL model learns to effectively parameterize the free-surface multiple-free wavefield from the full wavefield by incorporating the underlying physics into the loss computation. This, in turn, yields high-quality estimates without ever being shown any 'ground truth' data. Currently, the network reparameterization is performed independently for each dataset. We demonstrate its effectiveness through tests on both synthetic and field data. We employ industry-standard Surface-Related Multiple Elimination (SRME) using, respectively, global least-squares adaptive subtraction and local least-squares adaptive subtraction as benchmarks. The comparison shows that the proposed method outperforms the benchmarks in estimation accuracy, achieving the most complete primary estimation and the least multiple energy leakage, but at the cost of a higher computational burden.

**Introduction**

In seismic processing, primary estimation (the wavefield excluding only free-surface multiples, as defined in this paper) and multiple (free-surface multiple) elimination represent two complementary aspects of the same task: separating signals based on whether they have interacted with the free surface. This separation is a critical step in seismic data processing.

Over decades, the most widely used methods in the seismic industry includes Surface-Related Multiple Elimination (SRME), upon which many derivative methods, such as Closed-loop SRME (Lopez, 2016, Wang et al. 2017), which combines SRME with Estimated Primaries via Sparse Inversion (Groenestijn and Verschuur, 2009, Lin et al. 2013), have been developed. The foundational theory of SRME was first introduced by Verschuur et al. in 1992. From a geophysical perspective, multiples can be predicted by the convolution of the full wavefield.

The established practice of SRME methods consists of two parts: first, predicting the multiples, and second, subtracting the predicted multiples from the full wavefield. Effectiveness depends heavily on accurately matching the predicted multiples to the original full wavefield in energy level and event locations. Otherwise, insufficient removal of multiples or over/wrong subtraction, damaging the primaries, may occur. Based on different subtraction methods, examples of SRME-based methods include SRME with global least-squares adaptive subtraction (Global LS-SRME) and SRME with local least-squares adaptive subtraction (Local LS-SRME).

In recent years, Deep Learning (DL), due to its great performance in computer vision (CV), has attracted significant attention in our community. Many such methods rely on supervised learning, where user-defined labels are used to form input-output pairs for training. The deep neural network (DNN) is then optimized by minimizing the difference between its output and the ground truth using a selected loss criterion. However, ground truth data for primaries do not naturally exist in field-acquired seismic data. This represents a general dilemma when employing supervised learning in seismic processing tasks. To tackle this, common attempts include using results produced by traditional methods, such as SRME, as quality-compromised ground truth, which entails additional costs to run the traditional methods beforehand, or training on synthetic data and applying the trained DNN to real-field data which presents a difficult domain shift problem (Hou and Messud, 2021). Besides, in both cases, the DNN is subject to a potential learning limitation imposed by the imperfect/incomplete quality of the training data.

In this work, we propose a method described as a physics-driven self-supervised DL approach for estimating primaries, i.e., the wavefield without the free surface multiples. It directly integrates the governing equation describing the physical relationship between the full wavefield, with and without the influence of the free surface, into the DL framework. This results in the form of a novel physics-driven learning strategy. On this foundation, the DL model learns to predict the primary without ever being shown any ground truth examples during its learning process. We validate it on synthetic and real data, with Global LS-SRME and Local LS-SRME serving as benchmarks for comparison.

**SRME Theory**

In this paper, we use the hat symbol (^) to denote estimated variables (e.g., $\widehat{\mathbf{P}}$ represents the estimated full wavefield), and lowercase variables denote data in the time domain (e.g., $\mathbf{p}$ represents the full wavefield in the time domain). SRME predicts primary $\mathbf{P}_0$ by subtracting free-surface multiples $\mathbf{P}_0\mathbf{AP}$ from the full wavefield $\mathbf{P}$:

$$\mathbf{P}_0 = \mathbf{P} - \mathbf{P}_0\mathbf{AP}. \tag{1}$$

Uppercase variables represent matrices containing pre-stack data for one frequency. Here, the surface operator is defined as $\mathbf{A} = \mathbf{IRS}^{-1}$, where $\mathbf{I}$ is a unit matrix, $\mathbf{R}$ represents surface reflectivity, and $\mathbf{S}$ represents the source wavelet. A detailed explanation and formal derivation of SRME can be found in Verschuur et al. (1992). Equation (1) forms the basis of SRME and directly leads to:

$$\mathbf{P} = \mathbf{P}_0(\mathbf{I} + \mathbf{AP}) \tag{2}$$

Equation (2) is employed as the governing physical equation in our framewrok.

**The Proposed Physics-Driven Learning Method**

Figure 1 illustrates the framework of the proposed method. This paper addresses a 2D SRME problem, which requires fully dense sampling, meaning that ideally, the number of sources and receivers should be equal. Thus, although the input data $\mathbf{p}$ in Figure 1 is represented as a 3D cube, it still corresponds to a 2D SRME problem.

We employ a 3D convolutional neural network (CNN) consisting of one encoder, linked to the input full wavefield $\mathbf{p}$, and two decoders, linked respectively to the estimated primaries $\widehat{\mathbf{p}}_0$ and surface operator $\widehat{\mathbf{a}}$. The 3D nature of the CNN is designed to incorporate information across all dimensions, capturing both spatial and temporal dependencies in the data. The encoder extracts key features from the input, transforming it into a latent representation. Its basic blocks are composed of 3D convolutional layers, ReLU activation functions, and batch normalization. Instead of using pooling, which is a fixed operation, this encoder employs stride convolutions that enable learnable downsampling filters. The number of filters in the encoder layers increases progressively to 64, 128, 256, 512, and finally reaches 1024 in the latent space. The decoder that outputs the estimated primary mirrors the encoder's structure, employing 3D transposed convolutional layers to progressively construct the primaries from the latent space. The other decoder follows the same principle to progressively construct the surface operator but employs fully connected layers at the end, as it aims to estimate a 1D (spatially-invariant) filter.

In this training framework, the estimated primaries $\widehat{\mathbf{p}}_0$ and surface operator $\widehat{\mathbf{a}}$, both output by the 3D CNN, are not compared with ground-truth labels, unlike in a typical supervised learning approach. Instead, they are passed into the governing equation linking the full wavefield, primaries, and multiples, represented by a purple rectangle in Figure 1. This physical equation specifically refers to equation (2), which serves as the foundational principle for SRME. Its output, which can also be seen as the indirect output of the 3D CNN model, is expected to be a reconstruction of the full wavefield data $\hat{\mathbf{p}}$.

The difference between the original input $\mathbf{p}$ and the reconstructed full wavefield data $\hat{\mathbf{p}}$ is computed using a selected loss criterion, and backpropagation is used to iteratively update the parameters of the 3D CNN, including the weights of the encoder and decoders. The specific loss function employed in this work is the mean square error (MSE), which is defined as

$$J = \frac{1}{n_s}\sum_{i=1}^{n_s}\frac{1}{n_t \cdot n_r}\sum_{j=1}^{n_t}\sum_{k=1}^{n_r}\big(\mathbf{p}(i,j,k) - \hat{\mathbf{p}}(i,j,k)\big)^2 \qquad (3)$$

where $n_s$, $n_t$ and $n_r$ represent the number of sources, time samples and receivers of the input data.

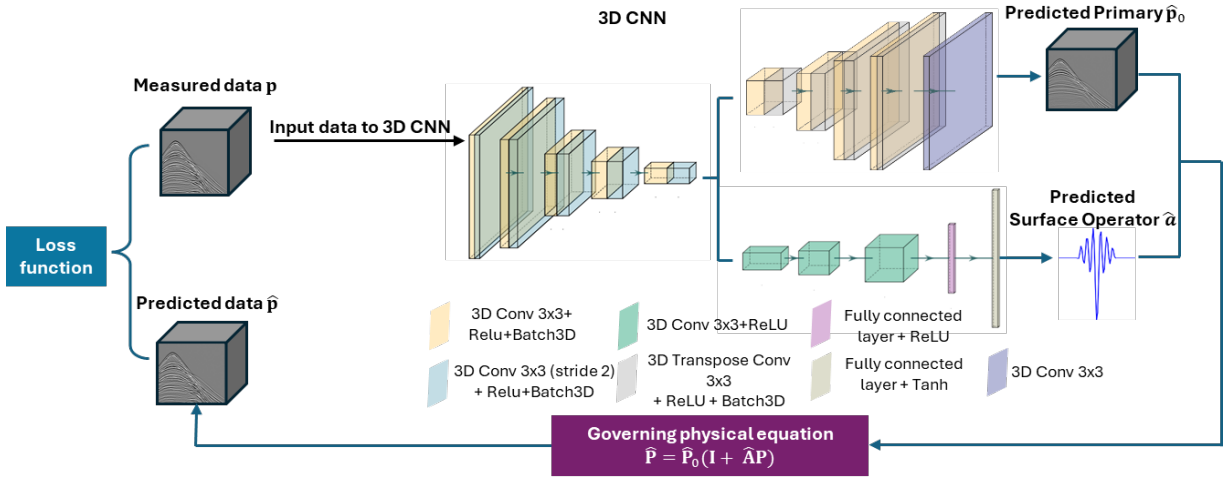

*Figure 1. The framework of the physics-driven self-supervised learning strategy with a visualization of the employed 3D CNN model. The governing physical equation, which links the full wavefield, primaries and multiples, serves as the foundation of the geophysical primary estimation method SRME. The input data $\mathbf{p}$ is a 3D cube representing the full wavefield, with dimensions corresponding to [shot, receiver, time]. The outputs are the estimated primaries $\hat{\mathbf{p}}_0$ and the surface operator $\hat{\mathbf{a}}$, where $\hat{\mathbf{p}}_0$ retains the same dimensions as the input data, and $\hat{\mathbf{a}}$ is a 1D signal representing a spatially invariant filter.*

### Synthetic Data Examples

To demonstrate the effectiveness of our proposed method, we first test it on synthetic data generated by the salt model (Lopez, 2016). The simulated survey includes 280 shots and 280 receivers, with a sampling interval of 8 ms and a lateral spacing of 15 m. An example of the simulated full wavefield data is shown in Figure 2a and the corresponding true primaries are given in Figure 2b. Global LS-SRME, Local LS-SRME are employed as benchmark methods, and it is worth noting that in both cases, the LS matching filters for SRME are derived in the shot domain; their primary estimation results are shown in Figures 2c and 2d, respectively. Finally, the estimated primaries from our proposed method are given in Figure 2e. The proposed method was trained for 998 epochs, with a total training time of 15 minutes on an H100 GPU.

Global LS-SRME is efficient to run; however, its primary estimation result (Figure 2c) contains noticeable multiple residuals throughout the entire sample. In contrast, Local LS-SRME achieves better accuracy, with most multiples attenuated and only minor residuals remaining in the deep region due to the overlap between primaries and multiples, as shown in Figure 2d. However, this improvement comes at the cost of requiring careful tuning of window parameters and filter lengths, which demands considerable effort to achieve optimal performance. Figure 2e presents the primary estimation result from our proposed method. Deep and shallow multiples are effectively suppressed, and the seismic events in the target zone (800–900 ms apex time) are notably clearer compared to the results from the benchmarks. Similar improvements are also evident in the far-offset region.

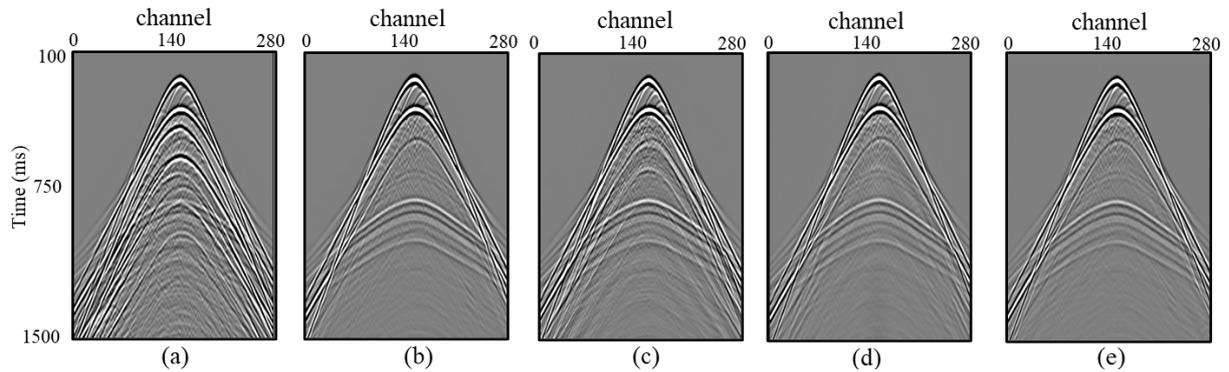

*Figure 2. Synthetic data examples: (a) full wavefield data; (b) true primaries; (c) to (f) estimated primaries from Global LS-SRME, Local LS-SRME, and the proposed method, respectively.*

**Field Data Example**

Further demonstration is conducted on 2D marine data (Lopez, 2016) consisting of 128 shots, each with 128 traces, and a sample interval of 8 ms. Both the shot and trace intervals are 25 m. The data has been pre-processed and regularized. An example of the input full wavefield data is shown in Figure 3a. As in the synthetic test, Global LS-SRME and Local LS-SRME are selected as benchmark methods. Their corresponding primary estimation results, along with the result from our proposed method, are shown in Figures 3b to 3d, respectively. The proposed method was trained for 681 epochs, with a total training time of 10 mins on an H100 GPU. The corresponding eliminated multiples are shown in Figures 3e to 3g. Note that surface multiples are expected to start only at 750 ms.

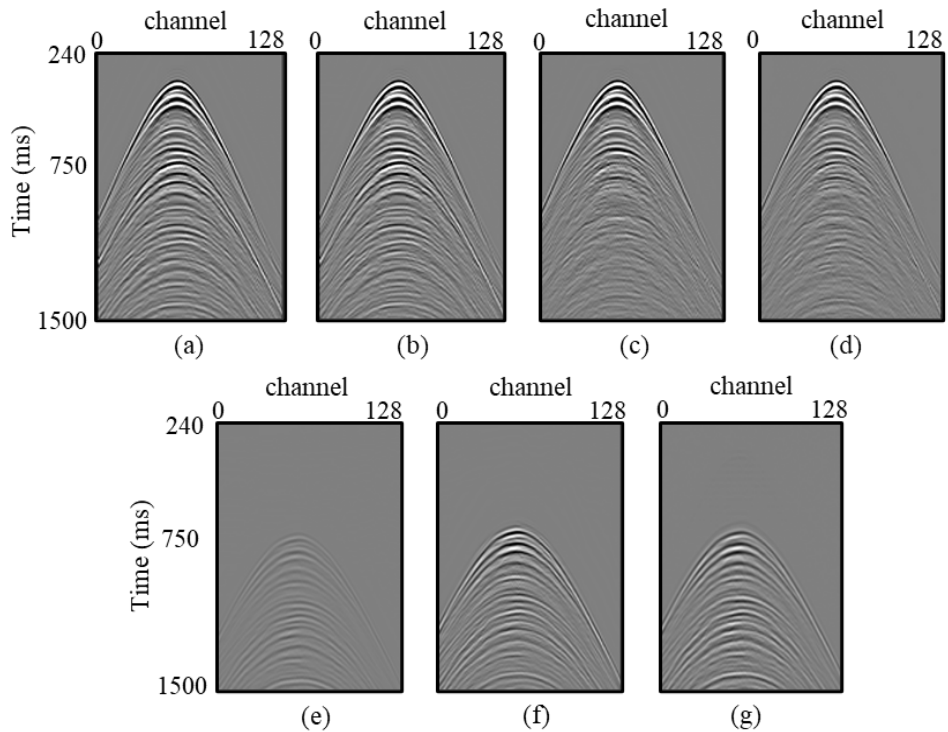

*Figure 3. Field data examples: (a) full wavefield data; (b) to (d) are the estimated primaries from Global LS-SRME, Local LS-SRME, and the proposed method, respectively; (e) to (g) are the corresponding eliminated multiples.*

As shown in Figures 3b and 3e, Global LS-SRME again leaves noticeable multiple residuals in the primary estimation result. Figures 3c and 3f show significant multiple elimination with very few primary leakages, indicating that after a proper parameter tuning, Local LS-SRME can achieve high effectiveness. In comparison, our proposed method provides comparably good multiple elimination to Local LS-SRME, while it promisingly achieves effective primary energy preservation, as shown in Figures 3d and 3g.

**Conclusions**

We propose a physics-driven self-supervised DL solution for free-surface multiple elimination. The key idea behind this framework is to embed the governing physics describing the relationship between the input wavefield and the desired output wavefield directly into the loss computation of the DNN. This approach transforms the model into a physics-driven learning engine, eliminating the need for large amounts of labeled training data. In this work, we apply this concept to primary estimation, a critical task in seismic processing. The framework's performance is evaluated through both synthetic and field data tests. The primary estimation results produced by the employed 3D CNN demonstrate the effectiveness of the proposed method, achieving higher accuracy compared to Global LS-SRME and Local LS-SRME.

**Discussions**

So far, the proposed method has been applied to each dataset independently, but there is a potential to extend it for simultaneous processing of multiple datasets, gnerateing different primary outputs in parallel. Additionally, the method may be generalized to learn a more flexible operator that can be applied across gathers in the same dataset and potentially to new datasets. Both of these extensions are still under investigation. We also acknowledge that computational efficiency could pose a challenge when addressing large-scale problems. The current work focuses on 2D scenarios, and the method was trained on an H100 GPU, 3D cases, with additional data scaling challenges, remain unexplored. Despite these limitations, the promising results observed in this study highlight the potential of the proposed framework for various other seismic data processing tasks, such as de-ghosting and de-blending.